\documentclass[12pt,a4paper]{article}
\usepackage{a4}
\usepackage{amssymb}
\usepackage{amscd}
\usepackage{amsmath}

\newcommand{\barre}[1]{%
        \setbox1=\hbox{$#1$} \dimen2=\ht1 \dimen3=\dp1 \dimen4=\wd1
        \setbox2=\hbox{\sl /}
        \dimen1=\wd1 \advance\dimen1 by -\wd2 \divide\dimen1 by 2
        \advance\dimen1 by \wd2 \advance\dimen1 by 0.4pt
        \setbox3=\hbox to \wd1{\hss \box1 \kern -\dimen1 \box2\hss}
        \ht3=\dimen2 \dp3=\dimen3 \wd3=\dimen4
        \box3
        }

\begin{document}


\begin{titlepage}
\begin{flushright}   BONN-TH-99-06  \end{flushright}
\vskip 2cm
\centerline{\LARGE{\bf {String Dualities in the Presence of}}}
\vskip .5cm
\centerline{\LARGE{\bf {Anomalous $U(1)$ Symmetries}}}
\vskip 2cm
\centerline{\bf Zygmunt Lalak${}^{\,}$\footnote{Permanent address:
Institute of Theoretical Physics, Warsaw University, Poland.},
St\'ephane Lavignac${}^{\,}$\footnote{Permanent address: Service
de Physique Th\'eorique, CEA-Saclay, F-91191 Gif-sur-Yvette C\'edex, France.}
and Hans Peter Nilles}
\vskip 1cm
\centerline{\em Physikalisches Institut, Universit\"at Bonn}
\centerline{\em Nussallee 12}
\centerline{\em D-53115 Bonn, Germany}

\vskip2cm
\centerline{\bf {Abstract}}
\indent

\vskip 1cm  

Anomalous U(1) gauge symmetries in
type II orientifold theories show some unexpected properties.
In contrast to the heterotic case, the masses of the gauge bosons
are in general of order of the string scale even in the absence of large
Fayet-Iliopoulos terms. Despite this fact, the notion of
heterotic-type II orientifold duality remains a useful concept,
although this symmetry does not seem to hold in all cases considered.
We analyse the status of this duality symmetry, clarify the
properties of anomalous U(1) gauge symmetry in the orientifold picture
and comment on the consequences for phenomenological applications
of such anomalous gauge symmetries.

\vfill
\end{titlepage}


\section{Introduction}

Since its appearance in 4-dimensional ($d=4$) heterotic string
theories, anomalous $U(1)$ gauge symmetries have recieved 
considerable attention and have been extensively used in model
building. One of the intriguing consequences of the presence of an
anomalous $U(1)$ in the heterotic theory
is the dynamical appearance of a
Fayet-Iliopoulos term $\xi$ in one-loop perturbation theory.
This induces nontrivial vacuum expectations values of charged
scalar fields that break the $U(1)$ and (potentially) other 
gauge symmetries spontaneously. This breakdown in connection
with the Green-Schwarz mechanism \cite{GS} renders the gauge
boson massive. In general the size of this mass and of $\xi$
is set by the string scale, possibly suppressed by a
constant factor $\epsilon \sim 10^{-2} - 10^{-1}$ that might have
interesting consequences for model building. All in all the
consequences of an anomalous $U(1)$ symmetry 
in the framework of the perturbative heterotic string are 
well understood.

The purpose of the present paper is to achieve a similar 
understanding of that situation in the framework of 
open string theories: the properties of anomalous $U(1)$
symmetries in type I and type II orientifolds. Here,
in contrast to the situation in the perturbative heterotic
theory, we might have to deal with several anomalous
$U(1)$ gauge symmetries \cite{Ibanez_orientifolds}
and a generalized Green-Schwarz mechanism \cite{Sagnotti_generalized,Berkooz}.
Originally it was assumed that each of these
symmetries comes with a one-loop Fayet-Iliopoulos term 
$\xi_i$ and that the Green-Schwarz mechanism involves all
the axions from the $d=10$ antisymmetric tensor multiplets,
including the ``model independent axion'' from the dilaton
supermultiplet. With this multitude of anomalous $U(1)$'s
one expected gauge boson masses and Fayet-Iliopoulos terms
$\xi_i$ of various sizes, including the possibility of
anomalous gauge boson masses that are small compared to the
string scale and might even tend to zero. In this paper we
want to show that these expectations are not justified and
try to clarify the situation. To do that, we rely on
two new results concerning the mechanism of anomaly
cancellation \cite{Ibanez_anomalous} and the generation of
Fayet-Iliopoulos terms \cite{Poppitz}.

We shall argue that in the orientifold picture, the values
of the $\xi_i$'s are moduli that can take arbitrary
values, but that the mass of the anomalous gauge bosons
nonetheless is large and independent of the $\xi_i$'s.
Thus the gauge boson masses are of the size
of the string scale even if some or all of the $\xi_i$'s
vanish. In that sense the role of the $\xi$'s here is
similar to those in a nonanomalous gauge symmetry where
it can usually be adjusted to any desired value. This is in
contrast to the heterotic theory, where we know that a
nonvanishing value of $\xi$ is induced. Of course,
one should be aware of the fact, that there might be
nonperturbative contributions to the $\xi$'s in the
orientifold picture. We shall come back to this question
later in the paper.

Although the situation is so 
different in the heterotic and open string theories, there
is the notion of heterotic-type I duality \cite{Polchinski}, 
that was assumed to hold in the orientifold picture as well
\cite{Sagnotti,Kakushadze_Z_3}. In the present paper we investigate
this duality in view of the results \cite{Ibanez_anomalous,Poppitz}
mentioned earlier. At the moment we are not able to give a general answer,
but we have to examine the situation on a model by
model basis. In some of the cases this duality
seems to hold exactly, although the role played by the dilaton
is different in heterotic and type I theories. Duality, where
it holds, leads to interesting results: a blown up orbifold on
the heterotic side can be dual to a type II model in the
exact orientifold limit (not blown up). We also confirm the
fact that anomalous gauge boson masses appear through the
Green-Schwarz mechanism even in the presence of vanishing
$\xi$'s, which are connected to the blowing up modes in
the orientifold case.

But this duality symmetry should be taken with a grain of salt. In
some of the cases (e.g. $Z_7$ and $Z_3\times Z_3$ orientifold)
problems appear at the level of maximally unbroken gauge
group or massless spectrum\footnote{ 
If we assume of sufficient breakdown of gauge groups,
as was done in previous investigations of duality
\cite{Kakushadze_Z_3,Kakushadze_Z_7,Kakushadze_Z_3xZ_3}, these
problems disappear, in the sense that states that are inconsistent
with duality are rendered heavy by the breakdown of gauge symmetries.}. 
At the moment we do not know how to interpret these
discrepancies. It could be used as an argument against the
validity  of duality, but it could as well be that we are
missing some nonperturbative mechanism that would restore it.
Such a mechanism could be a nonperturbative 
induction of Fayet-Iliopoulos terms in  type I theories.
This could lead to a further breakdown of gauge groups and a 
picture consistent with duality, but at the moment this
remains an open question.

The paper will be structured as follows. In the next chapter we
shall explain in detail properties of models with anomalous
$U(1)$ gauge theories. Chapter 3 will then give a discussion 
of the masses of anomalous gauge bosons. The question of
heterotic-type I duality will be analyzed in four examples
in chapter 4. The consequences for phenomenological application  
of anomalous $U(1)$ gauge symmetries will then be summarized
in chapter 5.


\section{The use of anomalous $U(1)$'s}

In field theoretic models we were taught to discard anomalous gauge
symmetries 
in order to avoid inconsistencies. This was even true for the
condition on the trace of the charges $\sum_i Q_i = 0$ of a
$U(1)$ gauge symmetry because of mixed gauge and gravitational
anomalies \cite{Alvarez}. Moreover a nonvanishing trace of
the $U(1)$ charges would reintroduce quadratic divergencies
in supersymmetric theories through a one-loop Fayet-Iliopoulos
term \cite{Fischler}. In string theory we then learned that
one can tolerate anomalous $U(1)$ gauge symmetries due to the
appearance of the Green-Schwarz mechanism \cite{GS} that 
provides a mass for the anomalous gauge boson. In fact,
anomalous $U(1)$ gauge symmetries are common in string
theories and could be useful for various reasons. Before we
discuss these applications in detail, let us first discuss the
appearance of the anomalous symmetries in various string models.

\subsection{$U(1)_A$ in heterotic string theory}

In this case one obtains models with at most one anomalous $U(1)$,
and the Green-Schwarz mechanism involves the so-called
model independent axion (the pseudoscalar of the dilaton
superfield $S$). The number of potentially anomalous gauge
bosons is in general limited by the number of antisymmetric tensor fields
in the ten-dimensional ($d=10$) string theory. This explains
the appearance of only one such gauge boson in the perturbative
heterotic string theory and leads to specific correlations between the
various (mixed) anomalies
\cite{Kobayashi}. This universal anomaly structure is tied
to the coupling of the dilaton multiplet to the various gauge
bosons.  

The appearance of a nonvanishing trace of the $U(1)$ charges leads
to the generation of a Fayet-Ilopoulos term $\xi^2$ at one loop.
In the low energy effective field theory this would be quadratically
divergent, but in string theory this divergence is cut off through
the inherent regularization due to modular invariance. One obtains
\cite{DSW,Atick}
\begin{equation}
  \xi^2\ \sim {1\over({S+S^*})}M_{\rm Planck}^2 \sim M_{\rm String}^2
\label{eq:FIterm}
\end{equation}
where $(S+S^*)\sim 1/g^2$ with the string coupling constant $g$.
The Fayet-Iliopoulos term of order of the string scale
$M_{\rm String}$ is thus generated in perturbation theory. 
This could in principle lead to a breakdown of supersymmetry,
but in all known cases there exists a supersymmetric minimum in
which charged scalar fields receive nonvanishing vacuum
expectation values (vevs), that break $U(1)_A$ (and even other
gauge groups) spontaneously. This then leads to a mixing of
the goldstone boson (as a member of a matter supermultiplet)
of this spontaneous breakdown and the
model-independent axion (as a member of the
dilaton multiplet) of the Green-Schwarz mechnism. 
One of the linear combinations will provide a mass to
the anomalous gauge boson. The other combination will obtain
a mass via nonperturbative effects that might even be 
related to an axion-solution of the strong CP-problem \cite{strong_CP}.

As we can see from (\ref{eq:FIterm}), both the mass of the
$U(1)_A$ gauge boson and the value of the Fayet-Iliopoulos term
$\xi$ are of the order of the string scale. Nonetheless,
models with 
an anomalous $U(1)$ have been considered under various
circumstances and lead to a number of desirable 
consequences. Among those are

\begin{itemize}

\item[(i)] the breakdown of some additional nonanomalous gauge groups
\cite{Font},

\item[(ii)] a mechanism to parametrize the fermion mass spectrum in
an economical way \cite{fermion_masses},

\item[(iii)]the possibility to induce a breakdown of supersymmetry
\cite{susy_breaking},

\item[(iv)] a satisfactory incorporation of D-term inflation \cite{inflation},

\item[(v)] the possibility for an axion solution of the strong
CP-problem \cite{strong_CP}.

\end{itemize}

The nice property of the perturbative heterotic string theory 
in the presence of an anomalous $U(1)$ is the 
fact that both $\xi$ and the mass of the anomalous gauge boson
are induced dynamically and not just put in by hand. Both of them,
though, are of order of the string scale $M_{\rm String}$, which
might be too high for some of the applications, notably (iv) and
(v). We will now compare this for the case of

\subsection{$U(1)_A$ in type I and type II orientifolds.}

These are in general $d=4$ string models of both open and
closed strings that are derived from either type I or type II
string theories in $d=10$ by appropriate orbifold or orientifold
projections \cite{orientifolds}. As a first surprise it was noticed, that in
these cases more than a single anomalous $U(1)$ symmetry could be obtained
\cite{Ibanez_orientifolds}. This led to the belief that here we can
deal with a new playground of various sizes of $\xi$'s and 
gauge boson masses in the phenomenological applications.

The appearance of several anomalous
$U(1)$'s is a consequence of the fact that these models contain
various antisymmetric tensor fields in the higher dimensional
theory and the presence of a generalized Green-Schwarz mechanism
\cite{Sagnotti_generalized,Berkooz} involving axion fields in new 
supermultiplets $M$. In the
type II orientifolds under consideration these new axion fields 
correspond to twisted fields in the Ramond-Ramond sector of the
theory.

From experience with the heterotic case it was then assumed
\cite{Kakushadze_Z_3} that for each anomalous $U(1)$ a 
Fayet-Iliopoulos term was induced dynamically. With a mixing of
the superfields $M$ and the dilaton superfield $S$ one 
hoped for $U(1)_A$ gauge boson masses of various sizes in 
connection with various sizes of the $\xi$'s.

The picture of duality between heterotic orbifolds and 
type II orientifolds as postulated in \cite{Sagnotti}
seemed to work even in the presence
of several anomalous $U(1)$
gauge bosons assuming the presence of Fayet-Iliopoulos terms
in perturbation theory and the presence of the generalized
Green-Schwarz mechanism.

Meanwhile we became aware of two decisive new results that
initiated our renewed interest in these questions and
forces us to reanalyse this situation. 
The first one concerns the inspection of the anomaly
cancellation mechanism in various type II orientifolds.
As was observed by Ib\'a\~nez et al. \cite{Ibanez_anomalous},
in this class of models there is no mixing between the dilaton
multiplet and the $M$-fields. It is solely the latter
that contribute to the anomaly cancellation. Thus the
dilaton that is at the origin of the Green-Schwarz mechanism
in the heterotic theory does not participate in that
mechanism in the dual orientifold picture.
The second new result concerns the appearance of
the Fayet-Iliopoulos terms. As was shown by Poppitz \cite{Poppitz}
in a specific model,
there were no $\xi$'s generated in one-loop perturbation theory.
The one loop contribution vanishes
because of tadpole cancellation in the given theory. 
This result seems to be of more general validity and
could have been anticipated from more general arguments,
since in type I theory a (one-loop) contribution to
a Fayet-Iliopoulos term either vanishes or is quadratically
divergent, and the latter divergence is avoided by the
requirement of tadpole cancellation. Of course, there is
a possibility to have tree level contributions to the $\xi$'s,
but they are undetermined, in contrast to the heterotic case
where $\xi$ is necessarily nonzero because of the
one loop contribution. In type II theory such a contribution
would have to be of nonperturbative origin.

In the heterotic theory the mass of the anomalous gauge boson was
proportional to the value of $\xi$. If a similar result would hold
in the orientifold picture, this would mean that some of
the $U(1)$ gauge bosons could become arbitrary light or even
massless, a situation somewhat unexpected from our experience
in quantum field theory. In any case a careful
reevaluation of several questions is necessary in the
light of this new situation. Among those are:

\begin{itemize}

\item the size of the $\xi$'s,

\item the size of the masses of anomalous $U(1)$ gauge bosons

\item relation of $\xi$ and gauge boson mass,

\item the fate of heterotic - type IIB orientifold duality,

\end{itemize}
which we will discuss in the remainder of this paper.


\section{Anomalous gauge boson masses}

\subsection{$D=4, N=1$ heterotic compactifications}

Let us first recall some facts about anomalous $U(1)$'s in $D=4, N=1$
compactifications of the heterotic string. The gauge group of such vacua
often possesses several abelian factors, one of which may be anomalous. Its
anomalies are harmless, however, since they are compensated for by a
four-dimensional version of the Green-Schwarz mechanism \cite{GS}
which ensures the consistency of the underlying $D=10$ string theory.
A Fayet-Iliopoulos term $\xi^2$ is generated in the $D=4$ vacuum at the
one string loop level, as well as a gauge boson mass at two loops. A string
computation gives \cite{Atick}
\begin{equation}
  \xi^2\ =\ \frac{\mbox{Tr}\, X}{192 \pi^2}\ M_{Str}^2
\label{eq:FI_het}
\end{equation}
where $\mbox{Tr}\, X$ denotes the trace of the anomalous charge, called
$X$ in the following, over all massless states of the theory.

As shown by Dine, Seiberg and Witten \cite{DSW}, much information about
the anomalous $U(1)$ can be obtained from a four-dimensional supersymmetric
formulation of the Green-Schwarz mechanism. Before going through this
formulation, let us recall how the Green-Schwarz mechanism
works in ten dimensions.
$D=10, N=1$ supergravity coupled to supersymmetric Yang-Mills theory has in
general gauge and gravitational anomalies which are generated by hexagon
diagrams with six external gauge bosons and/or gravitons. When the gauge
group is $SO(32)$ or $E_8 \times E_8\,$, all anomalies can be cancelled by
the addition of counterterms such as $B\, \mbox{tr}\, F^4\,$, where
$B$ and $F$ are the two-forms corresponding to the antisymmetric tensor
$B_{MN}$ from the supergravity multiplet and to the Yang-Mills field strength
$F_{MN}\,$, respectively ($M,N=0 \ldots 9$). This term, together with the
coupling $\partial^M B^{NP} \omega^{\scriptscriptstyle{YM}}_{MNP}$ present
in the supergravity action (where $\omega^{\scriptscriptstyle{YM}} =
\mbox{tr}\, (AF - \frac{1}{3} A^3)$ is the
Yang-Mills Chern-Simons three-form), generates an anomalous tree diagram
with a $B$ propagator and six external gauge fields. Such diagrams compensate
for the hexagon diagrams. It has been shown that the type I and heterotic
string theories automatically contain the counterterms required for anomaly
cancellation.

In $D=4, N=1$ heterotic vacua, abelian anomalies are compensated for by
a remnant of this mechanism.
The role of the ten-dimensional $B_{MN}$ is played by
the four-dimensional antisymmetric tensor $B_{\mu \nu}$ coming from the
components of $B_{MN}$ that are tangent to the non-compact dimensions. This
field couples to the four-dimensional Chern-Simons form:
\begin{equation}
  \partial^\mu B^{\nu \rho}\, \omega^{\scriptscriptstyle{Y\! M}}_{\mu \nu \rho}
\label{eq:B_w_het}
\end{equation}
and to the field strength of the anomalous $U(1)_X$:
\begin{equation}
  \epsilon_{\mu \nu \rho \sigma}\, B^{\mu \nu} F_X^{\rho \sigma}\ .
\label{eq:B_F_het}
\end{equation}
This last coupling is nothing but the four-dimentional remnant of the
ten-dimensio- nal Green-Schwarz counterterms (it can be obtained,
for example, by giving background expectation values to the field strengths
with compact indices in the term $B\, \mbox{tr}\, F^4$).

Let us now recall the standard supersymmetric formulation
of this mechanism \cite{DSW}. In four dimensions,
an antisymmetric tensor $B_{\mu \nu}$ describes only one degree of freedom and
is related through a duality transformation to a pseudo-scalar field:
$\partial_\mu B_{\nu \rho} \sim \epsilon_{\mu \nu \rho \sigma}\,
\partial^{\sigma} a\,$. After a duality transformation, the coupling
(\ref{eq:B_w_het}) may be rewritten (using the identity
$\mbox{d} \omega^{\scriptscriptstyle{YM}} = \mbox{tr}\, F^2$)
$a\, F^{\mu \nu} \widetilde{F}_{\mu \nu}$. Since the tree-level gauge kinetic
function of heterotic compactifications is simply the dilaton supermultiplet
(in the weakly coupled regime),
this tells us that the axion $a$ has to lie in this multiplet. Indeed, writing
$S \mid_{\theta = \bar \theta = 0}\ =\ s + i\, a\,$ and developping the gauge
kinetic terms in components, we obtain, omitting the fermionic terms:
\begin{eqnarray}
  {\cal L}_{GK} & = & \frac{1}{4}\: \sum_A \int \! \mbox{d}^2 \theta\
  S\, W^A W^A\ +\ \mbox{h.c.}  \nonumber \\
  & = & -\ \frac{1}{4}\ s\, F^{A \mu \nu} F^A_{\mu \nu}\
  +\ \frac{1}{4}\ a\, F^{A \mu \nu} \widetilde{F}^A_{\mu \nu}
  +\ \frac{1}{2}\: s\, D^A D^A
\label{eq:L_GK_het}
\end{eqnarray}
where the index $A = (a, i)$ runs over the factors of the four-dimensional
gauge group, $G = \otimes_a\, G_a\, \otimes_i U(1)_i$ (the $G_a$ are
simple groups, and $i=X$ corresponds to the anomalous $U(1)$), and we have
omitted the Kac-Moody levels for simplicity. In these
notations, the string coupling constant is given by $\langle s \rangle =
\frac{1}{g^2}\:$.

The Green-Schwarz counterterm (\ref{eq:B_F_het}), which after a duality
transformation becomes $\partial_\mu a\, A_X^\mu\,$, is described in a
supersymmetric manner by modifying the kinetic term of the dilaton,
$K (S, \bar S) = - \ln\, ( S + \bar S)\,$, to
\begin{equation}
  {\cal L}_K\ =\ - \int \! \mbox{d}^4 \theta\ \ln \left( S + \bar S
        - \delta\, V_X \right)
\label{eq:Kaehler_het}
\end{equation}
where the Green-Schwarz parameter $\delta$ characterizes the coupling of
the anomalous gauge boson to the axion. Gauge invariance then requires that,
under a $U(1)_X$ transformation with parameter $\Lambda_X\,$:
\begin{equation}
  S\ \rightarrow\ S\, +\, \frac{i}{2}\ \delta \Lambda_X
\label{eq:shift_axion}
\end{equation}
This results in a shift of the axion field $a \rightarrow a +
\frac{\delta}{2}\: \theta_X$ (where $\theta_X = \mbox{Re}\,\
\Lambda_X\mid_{\theta = \bar \theta = 0}$), which through the variation of
(\ref{eq:L_GK_het}) compensates for the anomaly\footnote{In (\ref{eq:anomaly}),
the anomaly coefficient $C_A$ is defined by $C_a = \sum_{R_a} 2\, T(R_a)
X_{R_a}$ for a non-abelian group $G_a$ (where $T(R_a)$ and $X_{R_a}$ are
respectively the index and the $X$-charge (\ref{eq:B_w_IIB}) of the
representation $R_a$), and by $C_i = 2\, \mbox{Tr}\, (Y^2_i X)$ for an abelian 
group $U(1)_{Y_i}$ (there is an additional symmetry factor for the cubic
anomaly of $U(1)_X\,$, $C_X = \frac{2}{3}\: \mbox{Tr}\, (X^3))$.}
\begin{equation}
  \delta {\cal L} \mid_{\mbox{1-loop}}\ =\ -\, \frac{\theta_X}{32 \pi^2}\
  \sum_A\, C_A\, F^A \widetilde{F}^A
\label{eq:anomaly}
\end{equation}
For this mechanism to work, the anomalies of the charge $X$ must satisfy the
relations:
\begin{equation}
  C_A\ =\ 4 \pi^2 \delta
\label{eq:anomalies_het}
\end{equation}
In heterotic compactifications, this is automatically the case. Thus all
anomalies (including the mixed gravitational anomaly $C_g = \mbox{Tr}\, X\,$,
which is also compensated for by the axion shift) are proportional to the
Green-Schwarz parameter $\delta\,$. This property, which ensures that heterotic
string vacua contain at most one anomalous $U(1)$, is a consequence of the
universal coupling of the dilaton superfield to the gauge fields. We shall
for this reason refer to the above anomaly compensation
mechanism as the {\it universal Green-Schwarz mechanism}.

Let us now develop the K\"ahler potential (\ref{eq:Kaehler_het}) in component
fields (again omitting fermionic terms):
\begin{equation}
  -\ \frac{1}{4\, s^2}\: \left( \partial^\mu s\,
    \partial_\mu s\, +\ \partial^\mu a\, \partial_\mu a \right)\
  +\ \frac{\delta}{4\, s^2}\: \partial_\mu a\, A_X^\mu\
  -\ \frac{\delta^2}{16\, s^2}\: A_X^ \mu A_{X \mu}\
  +\ \frac{\delta}{4\, s}\: D_X
  \label{eq:dilaton_Kaehler_components}
\end{equation}
where, since we are working in the Einstein frame, $M_{Pl}=1$ (here $M_{Pl}$
refers to the reduced Planck mass, $M_{Pl}=2.4 \times 10^{18} GeV$).
The supersymmetrization of the Green-Schwarz counterterm has introduced
a mass term for the anomalous gauge boson and a Fayet-Iliopoulos term. These
arise in string perturbation theory at the level of two loops and one
loop, respectively. Their expressions are given by: 
\begin{equation}
  M^2_X\ =\ \frac{1}{8}\ g^4 \delta^2 M^2_{Str}  \hskip 2cm
  \xi^2\ =\ \frac{\delta}{4}\: M^2_{Str}
\end{equation}
where we have restored the string scale $M_{Str} = g\, M_{Pl}\,$, and rescaled 
the vector multiplet $V_X \rightarrow g\, V_X$ in order to have canonical
kinetic terms for $A^X_{\mu}\,$. The Green-Schwarz parameter
$\delta$ is determined by the string computation of $\xi^2$ (\ref{eq:FI_het}):
\begin{equation}
  \delta\ =\ \frac{\mbox{Tr} X}{48 \pi^2}
\end{equation}
This fixes the proportionality coefficient between the mixed gravitational
anomaly and the gauge anomalies in (\ref{eq:anomalies_het}).

In the original vacuum, supersymmetry is broken by the Fayet-Iliopoulos term,
and the axion becomes the longitudinal component of the anomalous gauge boson.
In many compactifications, however, there exist shifted vacua in which
supersymmetry is preserved upon some scalar fields charged under $U(1)_X$
acquiring a vev: $\langle D_X \rangle = \sum_\alpha X_{\alpha}
|\langle \Phi_{\alpha} \rangle|^2 + \xi^2 = 0\,$. For this to happen, some of
the $X_{\alpha}$ must have the opposite sign to $\xi^2$. A combination of
the $S$ and the $\Phi_{\alpha}$ chiral superfields is then absorbed by the
anomalous vector multiplet and disappears from the
massless spectrum, while the orthogonal combination yields a low-energy
axion. The mass of the anomalous gauge boson is now
(assuming that $\xi^2$ is compensated for by a single vev with charge $X$)
$M^2_X = \frac{1}{8}\, g^2 \delta\, (g^2 \delta - 4\, X)\, M^2_{Str}\,$.
Note that, since the $\Phi_{\alpha}$ often carry other
charges, it is likely that other gauge groups, either abelian or non-abelian,
are broken together with the anomalous $U(1)$ \cite{Font}.

The universal Green-Schwarz mechanism then leads to the following picture:
once the dilaton assumes its vacuum expectation value, the anomalous $U(1)_X$
is broken and a Fayet-Iliopoulos term is generated. Comparing the two scales
\begin{equation}
  M^2_X\ =\ \frac{1}{2}\ g^2\, (g^2 \delta - 4\, X)\ \xi^2
\end{equation}
we see that while $\xi^2$ is tied to the string scale, one could in principle
make $M_X$ light with respect to $M_{Str}$ by lowering the string coupling
constant. This possibility would conflict, however, with the successful gauge
coupling unification of the MSSM.

\subsection{$D=4, N=1$ type IIB orientifolds}

As stressed in the introduction, $D=4, N=1$ type IIB orientifolds
show a very different
pattern of anomaly cancellation. The gauge group of such vacua may contain
more than one anomalous $U(1)\,$. Their anomalies are not universal in
the sense of Eq. (\ref{eq:anomalies_het}), and they are compensated by
a generalized version of the Green-Schwarz mechanism, which involves several
antisymmetric tensors. Also, a string computation \cite{Poppitz} has shown
that no Fayet-Iliopoulos term is generated at the one-loop level. While this
result has been obtained in a specific vacuum (the $Z_3$ orientifold of
Ref. \cite{Sagnotti}), it is believed to hold in a larger class of models,
since it is related to tadpole cancellation.

The cancellation of $U(1)$ anomalies in toroidal $\mathbb{Z}_N$ type IIB
orientifolds has
been studied in Ref. \cite{Ibanez_anomalous}. Let us summarize here their
results. In addition to the antisymmetric tensor $B_{\mu \nu}$ from the
untwisted sector, which is also present in heterotic compactifications, there
are several RR antisymmetric tensors $B_{k \mu \nu}, k=1 \ldots M$ from the
twisted sector, which are associated to the fixed points of the underlying
orbifold. Similarly to the heterotic $B_{\mu \nu}$, those twisted
antisymmetric tensors couple to the Yang-Mills Chern-Simons forms:
\begin{equation}
  \partial^\mu B_k^{\nu \rho}\, \omega^{A\, \scriptscriptstyle{Y\! M}}_{\mu
    \nu \rho}
\label{eq:B_w_IIB}
\end{equation}
and to the field strength of the $N$ abelian factors $U(1)_i$ present in the
gauge group:
\begin{equation}
  \epsilon_{\mu \nu \rho \sigma}\, B_k^{\mu \nu} F_i^{\rho \sigma}
\label{eq:B_F_IIB}
\end{equation}
But, contrary to the Green-Schwarz counterterm of heterotic compactifications,
the couplings (\ref{eq:B_F_IIB}) are present at tree-level. Another striking
difference with the heterotic case is that there is no such coupling for
$B_{\mu \nu}$ - implying that the dilaton superfield does not play any role
in anomaly cancellation.

The pseudoscalar duals $a_k$ of the twisted antisymmetric tensors lie in the
same chiral multiplets as the NS-NS twisted moduli $m_k$ corresponding to
the blowing-up modes associated with the singularities of the orbifold:
\begin{equation}
  M_k \mid_{\theta = \bar \theta = 0}\ =\ m_k + i\, a_k
\end{equation}
Performing a duality transformation on the couplings (\ref{eq:B_w_IIB}),
we obtain the following
expression for the gauge kinetic function:
\begin{equation}
  f_A\ =\ f_p\, +\, \sum_k\, c^k_A\, M_k
\label{eq:f_a_IIB}
\end{equation}
where $f_p$ is a function of the untwisted moduli\footnote{In type IIB
orientifolds, the dependence on the untwisted moduli of the gauge kinetic
function associated with a gauge group depends on the D-brane
sector this gauge group comes from. For example, gauge groups coming from
9-branes have $f_p = S$ \cite{Ibanez_orientifolds}.} 
and the $c^k_A$ are model-dependent coefficients. Similarly, the
couplings (\ref{eq:B_F_IIB}) can be rewritten:
\begin{equation}
  \sum_{i,\, k}\, \delta^k_i\, \partial_\mu a_k A_i^\mu
\end{equation}
where the Green-Schwarz parameters $\delta^k_i$ are model-dependent
coefficients as well\footnote{More precisely,
the $c^k_A$ and $\delta^k_i$ are given by
$c^k_A = \mbox{Tr}\, (\gamma_k^{-1} \lambda^2_A)$ ($A = a, i$) and
$\delta^k_i = \mbox{Tr}\, (\gamma_k \lambda_i)$ respectively, where $\gamma_k$
represents the action of the $k^{\mbox th}$ orbifold twist on the Chan-Paton
factors, and $\lambda_i$ ($\lambda_a$) is the Chan-Paton matrix associated
with the gauge group $U(1)_i$ ($G_a$) \cite{Ibanez_anomalous}.}.
As stressed before, however, the $\delta_i$ corresponding to the
model-independent axion always vanishes. Thus only the twisted moduli, and
not the dilaton, participate in the generalized
Green-Schwarz mechanism. This tells us that the K\"{a}hler potential for the
$M_k$ fields takes the generic form\footnote{For the sake of simplicity, we
are working in the basis where the kinetic terms for the twisted moduli are
canonical in the orbifold limit, namely
$\frac{\partial^2 K}{\partial M_k \partial M_l}\! \mid_{\scriptscriptstyle{M_k
= 0}}\ = \frac{1}{2}\, \delta_{kl}$.}:
\begin{equation}
  K\ =\ K\, \left( \{\: M_k + \bar M_k - 2\, \sum_{i=1}^{N}\,
    \delta^k_i\, V_i\: \}_{k=1 \ldots M} \right)
\label{eq:Kaehler_IIB}
\end{equation}
and that, under a $U(1)_i$ transformation with gauge parameter $\Lambda_i\,$,
the $M_k$ undergo a shift:
\begin{equation}
  M_k\ \rightarrow\ M_k\, +\, i\, \delta^k_i \Lambda_i
\end{equation}
while the dilaton, as well as the other untwisted moduli, remains
unshifted. Anomaly cancellation then requires the non-universal relations:
\begin{equation}
  C^i_A\ =\ 8\, \pi^2 \sum_k\: c^k_A\, \delta^k_i
\end{equation}

As mentioned above, a string computation \cite{Poppitz} has shown that no
Fayet-Iliopoulos term is generated at one-loop level in such orientifolds.
However, a tree-level $\xi^2$ can appear upon the $m_k$ assuming a vacuum
expectation value. As stressed in \cite{Ibanez_anomalous}, this statement
does not rely on any particular assumption regarding the K\"{a}hler potential.
In the presence of the Green-Schwarz counterterms $\partial_\mu a_k A^\mu_i\,$,
supersymmetry requires couplings
\begin{equation}
  -\, \sum_{i,\, k}\, \delta^k_i\, m_k D_i
\label{eq:m_D_coupling}
\end{equation}
which, if the orbifold singularities are blown up, generate Fayet-Iliopoulos
terms $\xi_i^2 = - \sum_k\, \delta^k_i\, \langle m_k \rangle\,$ (for generic
K\"ahler potentials, the $\xi^2_i$ receive other contributions than
(\ref{eq:m_D_coupling}), which also depend on the $\langle m_k \rangle$). It
should be stressed that, since the $m_k$ are moduli, the $\xi^2_i$ are
arbitrary. This is to be contrasted with the heterotic case, in which
$\xi^2$ is tied to the string scale by a model-dependent coefficient: in the
orientifold case, the Fayet-Iliopoulos terms are just moduli.

Since the values of the Fayet-Iliopoulos terms are arbitrary, one may wonder
whether this is also the case for the anomalous gauge boson masses - recall
that in the heterotic case,  $M^2_X$ is proportional to $\xi^2$. In
\cite{Poppitz}, it was noticed that $M^2_X$ has a string-scale value if the
K\"ahler potential for the twisted moduli (\ref{eq:Kaehler_IIB}) is a square.
Our goal here is to show that {\it this statement is of general validity
and does not depend strongly on the particular choice of the K\"ahler
potential}. We shall first consider the
orbifold limit $\langle m_k \rangle = 0\,$, in which the couplings
(\ref{eq:m_D_coupling}) are computed. In this case, one can
identify the scalar components of the massive vector multiplets as
combinations of the $m_k$ fields and compute their masses (which by
supersymmetry are the same as the gauge boson masses) without knowing the
K\"ahler potential explicitly. Indeed, the couplings (\ref{eq:m_D_coupling})
induce a mass matrix for the $m_k$ fields
\begin{equation}
  \mu^2_{kl}\ =\ \sum_i\, g^2_i\, \delta^k_i \delta^l_i
  \label{eq:moduli_matrix}
\end{equation}
(we have performed the canonical rescaling on the abelian vector
multiplets: $V_i \rightarrow g_i V_i\,$, where $g_i = \langle \mbox{Re}\,
f_i \rangle^{-1/2}$). This matrix is diagonalized by some rotation $R$:
\begin{equation}
  \mu^2_p\, \delta_{pq}\ =\ \sum_i\, g^2_i\, \bar \delta^p_i
  \bar \delta^q_i  \hskip 2cm
  \bar \delta^p_i\ =\ \sum_k\, R_{pk}\, \delta^k_i 
\label{eq:mp_masses}
\end{equation}
One can always choose $R$ such that the first $r$ eigenvalues are nonzero.
This defines in an unambigous way $r$ massive combinations of the $m_k$:
\begin{equation}
  m'_p\ =\ \sum_k\, R_{pk}\, m_k  \hskip 2cm  p\, =\, 1 \ldots r
\end{equation}
the $M-r$ remaining combinations, which can always be redefined, being
massless. Performing the same rotation $R$ on the NS-NS partners of the
scalars $m_k\,$, one finds that only the $a'_p\,$, $p=1 \ldots r$
couple to the abelian gauge bosons:
\begin{equation}
  \sum_{p=1}^r\, \mu_p\, \partial_{\mu} a'_p A^{\prime\, \mu}_p
\label{eq:mp_Ap_coupling}
\end{equation}
where we have redefined the $A^{\mu}_i$ to
\begin{equation}
  A^{\prime\, \mu}_p\ =\ \sum_i\, \frac{g_i \bar \delta^p_i}{\mu_p}\
  A^{\mu}_i  \hskip 2cm  p\, =\, 1 \ldots r
\label{eq:}
\end{equation}
and orthogonal combinations for the remaining $N-r$ gauge bosons
$A^{\prime \mu}_p\,$, $p=r+1 \ldots N\,$. Equations (\ref{eq:mp_masses})
and (\ref{eq:mp_Ap_coupling}) tell us that the chiral superfields
$M'_p = \sum_k R_{pk} M_k\,$, $p=1 \ldots r$ are absorbed by
the $r$ vector superfields $V'_p = \sum_i g_i \bar \delta^p_i
V_i\, / \mu_p$ to form $r$ massive vector multiplets. Then supersymmetry 
allows us to conclude that the abelian gauge bosons $A^{\prime \mu}_p$
have the same mass as the scalars $m'_p\,$.

We can ask the question how the above formulae are modified when one blows up
the orientifold. For a
generic K\"ahler potential, both the normalization of the kinetic terms of the
twisted moduli and their couplings to the abelian gauge bosons are corrected by
the non-vanishing of the blowing-up modes. These corrections can be taken into
account by moving to the basis where the kinetic terms of the twisted moduli
are canonical, $\widetilde M_k = \sqrt{2}\, \sum_l K^{1/2}_{kl}\, M_l$ (where
$K_{kl} = \partial^2 K / \partial M_k \partial M_l$ is the K\"ahler metric),
before applying the procedure of the previous paragraph. All formulae then
remain valid, with the $\delta^k_i$ replaced by moduli-dependent coefficients
$\widetilde \delta^k_i = \sqrt{2}\, \sum_l K^{1/2}_{kl}\, \delta^l_i\,$.
Therefore, in a blown-up orientifold, the gauge boson masses $\mu_p$
depend on the values of the blowing-up modes both through the gauge couplings
$g_i\,$, with $g^{-2}_i = \langle \mbox{Re} f_i \rangle = \langle \mbox{Re} f_p
\rangle + \sum_k c^k_i \langle m_k \rangle\,$, and through the K\"ahler
metric $K_{kl} = \sum_{N=0}^{\infty}
\frac{2^N}{N!}\, K_{kln_1 \ldots n_N}(0)\, \langle m_{n_1} \rangle
\ldots \langle m_{n_N} \rangle = \frac{1}{2}\,
\delta_{kl} + 2 \sum_n K_{kln}(0)\, \langle m_n \rangle + \ldots\,$.

It is not difficult to identify the combinations of the $U(1)$'s that
are anomaly-free. Eq. (\ref{eq:mp_Ap_coupling}) tells us that there are no
couplings between the $a'_p$ and the $N-r$ massless gauge bosons,
so the corresponding $U(1)$'s must be anomaly-free.
This can be checked explicitely by
redefining the charges accordingly to the gauge bosons, $Y'_p =
\sum_i R^V_{pi}\, g_i Y_i / \scriptstyle{\sqrt{\sum_i g^2_i}}\,$, where
$R^V_{pi} =  g_i \bar \delta^p_i / \mu_p$ for $p=1 \ldots r\,$, and by
computing the anomalies
in the new basis. One finds that the charges $Y'_{p=r+1 \ldots N}$ associated
with the massless gauge bosons have no anomalies, except
mixed anomalies with the anomalous charges $Y'_{p=1 \ldots r}$ which
are compensated for by the generalized Green-Schwarz mechanism. After
integrating out the massive vector multiplets, we end up with $N-r$
anomaly-free $U(1)$'s and $M-r$ twisted moduli $M'_{p=r+1 \ldots M}$.
Note that contrary to the heterotic case, a vacuum shift is not required to
maintain supersymmetry, unless the orientifold is blown up and
Fayet-Iliopoulos terms are generated. Eq. (\ref{eq:mp_Ap_coupling}) implies
that this can happen only for an anomalous vector multiplet $V'_p$, and only
if the scalar partner of the massive gauge boson $m'_p$ has a nonzero vev.
Indeed, the anomalous $D$-terms read:
\begin{equation}
  D'_p\ =\ -\,  \sqrt{{\scriptstyle \sum_i g^2_i}}\:
  \left( \sum_\alpha\, Y^{\prime\, \alpha}_p\, |\Phi_{\alpha}|^2 -\
    \frac{\mu_p}{\scriptstyle{\sqrt{\sum_i g^2_i}}}\ m'_p  \right)
  \hskip 2cm  p\, =\, 1 \ldots r
\end{equation}
At the level of unbroken supersymmetry, and in the absence of any
nonperturbative mechanism that would stabilize them, the vevs of the $m'_p$
are only restricted by the vanishing of the anomalous
$D$-terms (together with the vevs of the matter fields $\Phi_{\alpha}\,$,
which are constrained by the other $D$-terms as well). Thus nothing forces them
to be nonzero, and the Fayet-Iliopoulos terms in type
IIB orientifolds are just moduli. In particular, there is an obvious vacuum
$\langle \Phi_{\alpha} \rangle = 0\,$, $\langle m_k \rangle = 0\,$,
corresponding to the orbifold limit, in which all $\xi^2_p$ vanish. In this
vacuum, only the anomalous $U(1)$'s are broken, and their associated gauge
bosons become heavy and decouple, leaving $r$ residual global symmetries.
On the other hand, any nonzero $\langle m'_p \rangle\,$,
$p=1 \ldots r$ would force some of the matter fields $\Phi_{\alpha}$ to
acquire a vev, possibly leading to further breakdown of the gauge group.

To summarize, in $D=4, N=1$ type IIB orientifolds, the vector supermultiplets
$V'_p$ associated with the anomalous $U(1)$'s become massive by
absorbing a twisted modulus chiral supermultiplet $M'_p$. The masses
of the massive multiplets can be computed from the diagonalization of the
twisted moduli mass matrix (\ref{eq:moduli_matrix}):
\begin{equation}
  \mu^2_p\ =\ \sum_i\, g_i^2\, \bar \delta^p_i \bar \delta^p_i\,
  M^2_{Pl}  \hskip 2cm  p\, =\, 1 \ldots r
\label{eq:V_masses}
\end{equation}
(where we have restored the Planck mass). To each of these massive
multiplets is associated a moduli-dependent Fayet-Iliopoulos term
\begin{equation}
  \xi^2_p\ =\ -\, \frac{\mu_p}{\scriptstyle{\sqrt{\sum_i g^2_i}}}\
  \langle m'_p \rangle\ M_{Pl}  \hskip 2cm  p\, =\, 1 \ldots r
\label{eq:relation_xi_mu}
\end{equation}
which is proportional to the vev of the scalar partner of the gauge boson.
From the relation (\ref{eq:relation_xi_mu}) we conclude that the anomalous
gauge boson masses and their Fayet-Iliopoulos terms are essentially decoupled.
Indeed, the latter can be made arbitrarily small by tuning the
$\langle m_k \rangle\,$, while the former have a Planck-scale value in
the orientifold limit, from which they can possibly depart only for large
values of the blowing-up modes.
This is to be contrasted with the heterotic case, in which the relation
between the Fayet-Iliopoulos term and the anomalous gauge boson mass is
controlled by the gauge coupling.

One may still ask whether it is possible to make the anomalous gauge bosons
light for large values of the blowing-up modes. Let us consider for simplicity
the case where the gauge group
contains a single abelian factor $U(1)_X$ (this is the case in the $Z_3$
orbifold \cite{Sagnotti}), with gauge coupling $g_X\,$. The gauge boson mass
and the Fayet-Iliopoulos term are, respectively:
\begin{equation}
  M^2_X\ =\ 2\, g^2_X \sum_{k,l}\, K_{kl}\, \delta_X^k \delta_X^l\, M^2_{Pl}
  \hskip 1cm  \xi^2\ =\ -\, 2 \sum_{k,l}\, K_{kl}\, \delta^k_X
  \langle m_l \rangle\, M^2_{Pl}
\end{equation}
Since $g^{-2}_X = \langle \mbox{Re} f_X \rangle = \langle \mbox{Re} f_p
\rangle + \sum_k c^k_X \langle m_k \rangle$, one could try to make $M^2_X$
small by giving very large values to the blowing-up modes.
However, if the K\"ahler potential differs from a square, the large values of
$\langle m_k \rangle$ also contribute to the K\"ahler metric, $K_{kl} =
\frac{1}{2}\, \delta_{kl} + 2 \sum_n K_{kln}(0)\, \langle m_n \rangle
+ \ldots\,$, making it less natural to envisage a light anomalous gauge boson.
Even if the K\"ahler potential were quadratic, a nonzero value of the
$m_k$ would induce a vev of some field with anomalous charge $X$, resulting in:
\begin{equation}
  M^2_X\ =\ \frac{\sum_k \delta_X^k \delta_X^k + 2 X \sum_k
    \delta^k_X \langle m_k \rangle}{\langle \mbox{Re} f_p \rangle + \sum_k
    c^k_X \langle m_k \rangle}\ M^2_{Pl}
\end{equation}
Thus it would be very difficult, even in this case, to obtain light gauge
bosons. To conclude, one does not expect the blowing-up of the orientifold to
lower the masses of the anomalous gauge bosons. The safest possibility to
make them light would be to tune the values of the untwisted moduli so as
to make the gauge coupling small, much like in the heterotic
case.

\section{Heterotic-Type I Duality}

The class of models containing anomalous $U(1)$ factors offers a
playground for studying details of Type I - Heterotic duality in
four dimensions. As pointed out in \cite{Sagnotti} this duality,
which is of the weak coupling - strong coupling type in ten
dimensions, upon compactification to lower dimensions gives rise
to weak coupling - weak coupling dualities in certain portions of
the moduli space. In four dimensions the relation between the
heterotic and type I dilatons is
\begin{equation}
\phi_H = \frac{1}{2} \phi_I - \frac{1}{8} \log (G_I)
\end{equation}
where $G_I$ is the determinant of the metric of the compact 6d
space, which depends on some of the moduli fields. As the string
coupling is $e^{\phi_{I,H}}$, and the volume of the compact space
is at least unity in string units for phenomenologically relevant models,
then it is obvious that dual models on both sides
can be simultaneously weakly coupled. One may question whether this
requirement of weak string coupling on both sides constrains in
any way the values of the moduli which we need to solve D-and
F-flatness conditions as well as to give masses to unwanted
particle states. The answer is no and does not depend on the model.
The basic observation is
that the generalized Green-Schwarz terms which we find on type I
side do not depend on the dilaton (i.e. on the 4d universal S
modulus), and the compact space volume does not depend on the
twisted moduli $M$ which enter the generalized Green-Schwarz terms
and, consequently, four dimensional anomalous D-terms. The
independence of the anomalous $U(1)$ D-terms of the dilaton does not
hold on the heterotic side, where the Fayet-Iliopoulos term
$\xi_H$ depends on the dilaton only, and not on any other modulus.
But we know already from earlier sections, and shall see in more
detail here, that thanks to the existence of certain states
charged under anomalous $U(1)$, these models also fulfill all
the consistency checks, like the requirement of the unification of
gauge couplings at the proper scale and value.

Before we procede to analyze specific examples, let us specify
the criteria for two models to be called dual to each other.
First of all, we remind the reader that already in the
heterotic models the anomalous $U(1)$ appears in the low energy
lagrangian only to cause a shift of the vacuum which restores
supersymmetry, and to be immediately decoupled in a supersymmetric
manner. What is left behind, is the supersymmetric model with a
global $U(1)$ symmetry realized on the matter supermultiplets in a
linear way (the moduli which remain massless do not transform
under the global $U(1)$). Thus, already there, the perturbative
couplings between would-be `light' states play an important role
in finding the correct supersymmetric vacuum. The same phenomenon
is found in the present case on both sides, and, moreover, to
establish the duality equivalence of two models, we shall need
perturbative superpotential couplings
 between light fields on the heterotic side.
Hence, we shall be working at the level of the effective
lagrangian valid just below the respective string scale on each
side, heterotic and type I \footnote{The scales and couplings in
orientifold models were recently discussed in \cite{Ibanez_aspects}.}.
For the duality to hold between two models we require that they
have supersymmetric families of vacua, and that the spectrum of
the massless excitations around these vacua is equivalent. This
means, in particular, that the unbroken gauge groups and their
massless representations should be the same. In the sector of
gauge singlet fields, we require that the number of truly massless
states be the same on both, heterotic and type I, sides. In
addition, we require that the masses of the states which become
massive upon the choice of the vacua of the field theoretical
lagrangians we are analyzing be of the same order of magnitude.
This requirement means in particular that we expect the masses of
the gauge bosons of the anomalous $U(1)$s to be very close to each
other among the dual pairs of models. In practice, we solve the
D-flatness conditions for all anomalous and nonanomalous $U(1)$
groups on both sides
and then for the F-flatness conditions on the heterotic side.
It is legitimate in the string context to assume that our unbroken
supersymmetry minimum corresponds to flat space and, hence, that
the use of the globally supersymmetric lagrangian instead of the
locally supersymmetric one is justified.

The pairs of models which we study are type IIB orientifolds models in 4d
and their candidate heterotic duals which can be found in the existing
literature
\cite{Sagnotti,Kakushadze_Z_3,Kakushadze_Z_7,Kakushadze_Z_3xZ_3,Ibanez_Z_3,
Ibanez_orientifolds,Kakushadze_orientifolds,Lykken}.

\subsection{$Z_3$ models without Wilson lines}

The first two examples are $Z_3$ orientifolds/orbifolds without and
with Wilson lines. The model without Wilson lines is actually
the original example proposed for the conjectured type I - heterotic
duality in four dimensions in \cite{Sagnotti}. The type IIB
orientifold model has the gauge group $G=SU(12) \times SO(8)
\times U(1)_A$ where the $U(1)_A$ factor is anomalous. The
anomalies are non-universal and get cancelled by means of the
generalized GS mechanism. This mechanism involves twenty-seven
twisted singlets $M_{\alpha \beta \gamma}$, a particular
combination of which combines with the anomalous vector
superfield to form a massive multiplet. After the
decoupling of this heavy vector multiplet we obtain the
nonanomalous model with the gauge goup $G'=SU(12) \times SO(8)$.

On the heterotic side, which is the heterotic $SO(32)$ superstring
compactified on $T^6/Z_3$, the gauge group is $G=SU(12)
\times SO(8) \times U(1)_A$ and the $U(1)_A$ is again anomalous.
Its anomalies, however, are universal in this case, and the
universal, only dilaton-dependent, Fayet-Iliopoulos parameter is
generated. In this case there are also fields which are charged only
under the anomalous $U(1)$ and that can compensate for the Fayet-Iliopoulos
term by assuming an expectation value, whithout breaking the gauge group
any further; a combination of these fields and of the dilaton supermultiplet
is absorbed by the anomalous vector multiplet. These nonabelian singlets
are the counterparts of the
$M_{\alpha \beta \gamma}$ moduli of the orientifold model. However, on
the heterotic side we have additional states charged under $U(1)_A$ (and
also under $SO(8)$) the counterparts of which are not present in
the orientifold model. These unwanted states become heavy in a supersymmetric
manner through the superpotential couplings \cite{Kakushadze_Z_3}
\begin{equation}
W_H = \Lambda_{\alpha \beta \gamma \; \alpha' \beta' \gamma' \; \alpha''
\beta'' \gamma''}
Tr ( M_{\alpha \beta \gamma } V_{\alpha' \beta' \gamma' } V_{\alpha'' \beta''
\gamma'' })\ .
\end{equation}
Upon giving expectation values to the $M$'s, the supermultiplets $V$ obtain
supersymmetric mass terms of the order of $\xi$.
Below the scale of the heavy gauge boson mass we have a pair of
models which exactly fullfills our duality criteria.

One should note that on the heterotic side we have a blown-up orbifold,
since the scalars that
assume a vacuum expectation value correspond to the blowing-up modes.
Thus, in this case, a Type IIB orientifold is found to be dual to a blown-up
heterotic orbifold\footnote{The blowing-up of the $Z_3$ orientifold has
been recently discussed in Ref. \cite{Cvetic}.}.
The next point to be stressed is that the duality works
even though no Fayet-Iliopoulos term is present on the orientifold side.
In Ref. \cite{Kakushadze_Z_3} where, according to the general belief,
the generation of a 1-loop Fayet-Iliopoulos term in the orientifold model
had been assumed, duality held only in a region of the moduli space where the
nonabelian gauge groups are broken. If such a term were generated on the
Type IIB side, perhaps by a
nonperturbative mechanism, the duality would still hold, but one
would have to blow up the orientifold to achieve D-flatness on the
Type IIB side.

\subsection{$Z_3$ models with a discrete Wilson line}

One can add a discrete Wilson line to the $Z_3$ orientifold
construction \cite{Ibanez_Z_3}. In this case the gauge group of the
orientifold model is $G=SU(4)^4 \times U(1)^4$ where three of the
four  $U(1)$s are anomalous and decouple, as they become massive upon
mixing with three combinations of the orientifold blowing-up
modes. The combination of the four $U(1)$ generators which is
orthogonal to the three anomalous generators defines a
nonanomalous $U(1)$ under which all fields in the massless
spectrum are neutral. Hence, this last abelian factor cannot be spontaneously
broken, and the associated abelian vector boson remains in the massless
spectrum of the orientifold model. On the heterotic side there is, as always,
only one anomalous $U(1)_A$ which gets massive through the universal GS
mechanism. The fields which participate in forming the massive
vector multiplet are the dilaton and 27 fields which are neutral
under the nonabelian factors but have nonzero abelian charges: $9
\, (1,1,1,1)_{ \frac{4}{3}, \frac{8}{3}, 0, 0 } \, \oplus \, 9 \,
(1,1,1,1)_{ \frac{4}{3}, -\frac{4}{3}, -4, 0 } \, \oplus \, 9 \,
(1,1,1,1)_{ \frac{4}{3}, -\frac{4}{3}, 4, 0 }$ , where the subscript
numbers are the abelian charges (the anomalous charge is the first one).
Giving vevs to three combination of these fields allows to
make all abelian D-terms vanish, while breaking spontaneously the first two
nonanomalous $U(1)$ factors together with the anomalous $U(1)$.
The last $U(1)$ is not broken, as all the fields given above are neutral
with respect to it. Below the scale of breaking of the three $U(1)$'s,
which lies slightly below the string scale, the corresponding
gauge bosons decouple, and both models have
the same gauge groups $G'=SU(4)^4 \times U(1)$ and the same
massless spectra at the end. It should be noted that this example
contains on the Type IIB side three independent anomalous $U(1)$
factors, which is a real novelty when compared to the heterotic
models where one gets always a single anomalous $U(1)$.

\subsection{$Z_7$ models}

There exist, however, examples where exact duality (in the
sense specified at the beginning of this section) cannot be
achieved. The first of the examples we present here is the $Z_7$
orientifold/orbifold model given in \cite{Kakushadze_Z_7}. The orientifold
model has the gauge group $G=SU(4)^3 \times SO(8) \times U(1)^3$.
All three $U(1)$ factors are anomalous and their gauge bosons
decouple upon getting masses by the nonuniversal GS mechanism.
These gauge bosons mix with combinations of the chiral superfields
$M^{a}_\alpha$ which transform nonlinearly under the $U(1)$'s. In
this case the unbroken gauge group is large, $G'=SU(4)^3 \times
SO(8)$, since the inspection of the D- and F-flatness conditions
shows that the charged fields are not forced to assume vevs
breaking the nonabelian subgroups. The situation is very different
on the heterotic orbifold side. Here we have a unique
anomalous $U(1)$ and a Fayet-Iliopoulos term $\xi^2 \propto TrQ \,
> \, 0$. The only fields at hand which can cancel the anomalous
D-term and participate in giving a mass to the gauge boson are the
$Q$'s from the table given below (where the indices give $U(1)$ charges
and the first abelian factor is anomalous).
\begin{center}
\begin{tabular}{|c|c|c|}
\hline \hline
 &$ Q_1 $& $(1, \bar{4}, 4, 1)_{-2,1,1}$ \\
 untwisted&$ Q_2 $& $(\bar{4}, 1,4,1)_{-2,1,-1} $\\
 (partial)&$ Q_3 $& $(\bar{4}, \bar{4},1,1)_{-3,-3,0}$\\ \hline \hline
 &$ M^{1}_{\alpha} $& $7 \,
 (1,1,1,1)_{0,\frac{36}{7},-\frac{6}{7}}$ \\
twisted&$ M^{2}_{\alpha} $& $7 \,
 (1,1,1,1)_{0,-\frac{12}{7},\frac{20}{7}}$ \\
$\alpha =1 \ldots 7$&$ M^{3}_{\alpha} $& $7 \,
 (1,1,1,1)_{0,-\frac{24}{7},-\frac{16}{7}}$ \\ \hline \hline
&$ V^{1}_{\alpha} $& $7 \,
 (1,1,6,1)_{2,\frac{8}{7},-\frac{4}{7}}$ \\
twisted&$ V^{2}_{\alpha} $& $7 \,
 (1,6,1,1)_{2,\frac{2}{7},\frac{6}{7}}$ \\
$\alpha=1 \ldots 7$&$ V^{3}_{\alpha} $& $7 \,
 (6,1,1,1)_{0,-\frac{10}{7},-\frac{2}{7}}$ \\
\hline \hline
\end{tabular}
\end{center}
Since those fields are charged under the $SU(4)^3$ nonabelian factor, this
group is spontaneously broken together with the
nonanomalous $U(1)$ at the string scale,
and the low-energy gauge group is different
from that on the Type IIB side. The second problematic aspect is
that the fields $M^{a}_\alpha$ of the heterotic model must acquire vevs in
order to make massive in a supersymmetric way the unwanted states
$V^{a}_\alpha$, which are not present in the orientifold model.
However, on the orientifold side the corresponding
$M^{a}_\alpha$ states are gauge singlets,
and nothing forces them to assume nonzero vacuum expectation values.

Thus, in the $Z_7$ example neither the low energy gauge groups nor
the massless spectra match in the supposedly dual pair, at least
at the level of the perturbative effective lagrangian we rely on
here. The question is whether a nonperturbative contribution to
the superpotential or, perhaps a nontrivial K\"ahler potential
dependence on the fields $M^{a}_\alpha$ would change the picture.
The second type of corrections, although somewhat exotic in details,
could achieve duality. This comes from the fact that
certain additional contributions to the K\"ahler potential would
enforce nonzero vevs for the $M^{a}_\alpha$ states on the Type IIB
side (through the D-flatness conditions) and then the two models
could appear as a dual pair. The same effect would be
achieved if nonzero Fayet-Iliopoulos terms were generated, perhaps by
nonperturbative effects.

\subsection{$Z_3 \times Z_3$ models}

A second example which sheds doubts on the exact weak-weak
coupling 4d duality conjecture is the $Z_3 \times Z_3$
orbifold/orientifold constructions of Ref. \cite{Kakushadze_Z_3xZ_3}.
In this case
after giving masses to the anomalous gauge bosons (there is just
one on each side) the gauge group is the same in both models in
the pair, namely $G'=SU(4)^3 \times SO(8) \times U(1)_1 \times
U(1)_2$, but the spectra cannot be matched. There are massless
states in the heterotic model which are charged under the
nonanomalous $U(1)_1 \times U(1)_2$, which is not the case on the
orientifold side - the corresponding massless states are
neutral under the $U(1)_1 \times U(1)_2$ factor.

The direct inspection shows that in this particular case no
obvious modification of the K\"ahler potential, or nonperturbative
superpotential, can help restoring duality.

\subsection{Global anomalous $U(1)$ symmetries}

As pointed out long ago by Witten \cite{Witten}, when the
`anomalous' gauge boson decouples, there remain many chiral
superfields in the massless spectrum which were charged under that
$U(1)$, and their interactions, in particular the perturbative
superpotential, still respect the global version of that symmetry.
Under this global $U(1)'$ chiral fields transform linearly with
inherited charges, hence this global symmetry is anomalous. The
low energy dilaton, $S'$, does not transform under this symmetry.
The reason is that in terms of the original string variables the
chiral multiplet which is absorbed by the gauge boson to form the
massive vector multiplet is a linear combination of the original
dilaton, $S$, and the charged chiral multiplets which obtain the
vacuum expectation values. The composition of the combination
which is eaten depends on the vacuum configuration, but it always
contains an admixture of $S$. The orthogonal combinations are
massless, and one of them, $S'$, enters the gauge fields kinetic
functions the way the original dilaton did: $(S'\, W^\alpha
W_\alpha)_F + \,h.c.\,$. This combination does not have any other
couplings, hence it supports another anomalous global $U(1)''$
symmetry under which $S' \longrightarrow S' + i \, \gamma$ and
other fields are inert\footnote{This acts like the global shift of
the model independent axion in original string variables.}. There
is a unique combination of the $U(1)'$ and $U(1)''$ which is
anomaly free, and the orthogonal combination which is anomalous.
This, somewhat simplified but physically accurate, reasoning
convinces us that there is a global anomalous $U(1)$ symmetry left
behind the original anomalous local $U(1)$. This is the statement
in the context of the heterotic string compactifications, where
there is always only one local anomalous $U(1)$. Hence, on the
heterotic side we expect naturally one anomalous global $U(1)$
symmetry. Then, the question is what happens in the Type IIB
orientifold models, where one has more anomalous $U(1)$ factors,
like for example in the $Z_3$ with Wilson lines, or in $Z_7$
models discussed earlier. The answer is that the phenomenon
described above occurs seperately for each anomalous factor in
exactly the same way as described above. The role of the dilaton is
played this time by the nonuniversal fields $M$ which are numerous
in these models and whose combinations help to form massive vector
supermultiplets on the Type IIB side. Among the orthogonal
combinations are the superfields $M'$ which are completely
analogous to the field $S'$. At this point we can illustrate this
mechanism in the Type II case with a simple example. Let us take
for simplicity a single modulus $M$ and a single charged chiral
field $Y$. The K\"ahler potential with the two fields is
$K=\frac{1}{2}(M+\bar{M} -2 \delta V)^2 + Y \bar{Y}$, and the
relevant kinetic function $f=S + M$, where $g^2 = Re(f)^{-1}$. The
Lagrangian is \begin{equation} {\cal L} = \partial M \partial
\bar{M} +
\partial Y
\partial \bar{Y} - \frac{1}{2} g^2 ( M + \bar{M} - Y \bar{Y} )^2
\end{equation}
 Let us take the vacuum expectation values along the real
directions of the fields: $m=Re(M),\;y=Re(Y)\;$. Then one can find
the eigenvalues of the mass matrix of fluctuations around the
vacuum given by $\langle m \rangle, \; \langle y \rangle$, and
corresponding eigenmodes. The zero eigenmode is \begin{equation}
\phi_0 = \frac{1}{\sqrt{1 + \langle y \rangle^2}} (\langle y
\rangle \delta m + \delta y ) \end{equation}  and the orthogonal
mode with the mass $m^2=4 g^2( 1 + \langle y \rangle^2 )$ is
\begin{equation}  \phi_m = \frac{1}{\sqrt{1 + \langle y
\rangle^2}} (\langle y \rangle \delta y - \delta m )
\end{equation}  One can express the gauge kinetic function in
terms of the eigenmodes which we have found \begin{equation} Re(f)
= Re(S) +\langle m \rangle + \frac{1}{\sqrt{1 + \langle y
\rangle^2}} (\langle y \rangle \phi_0 - \phi_m ) \end{equation}
The heavy mode $\phi_m$ becomes part of the heavy gauge boson
multiplet and decouples from the massless fields. The field which
is a flat direction of the potential and enters the gauge kinetic
function is $\phi_0$. To make the expression for the effective
gauge kinetic function more transparent, it is convenient to
define the field $M'$ through $ Re(M') =\langle m \rangle +
\frac{1}{\sqrt{1 + \langle y \rangle^2}} \langle y \rangle
\phi_0$. Then the gauge kinetic funtion is simply $f_{eff} = S +
M'$. We have assumed that the heavy mode $\phi_m$, together with
its whole supermultiplet decouples completely from the massless
fields (which is precisely the case for the models discussed
here). Then the anomalous global symmetry acting on massless
fields is unbroken at the renormalizable level. However, one
should bear in mind that when we consider the full set of fields,
light and heavy, then the anomalous global $U(1)$ is spontaneously
broken. This means, strictly speaking, that even in the sector of
massless fields this symmetry shall be broken through suppressed
interactions with the heavy fields. We are working here with the
renormalizable interactions only, hence we can justifiably treat
the global symmetries as exact ones.

The same mechanism works for each anomalous $U(1)$ factor. This,
however, leads to the conclusion that in Type IIB models we have
several global
anomalous $U(1)$ symmetries. This conclusion is correct, as can
easily be verified for example in the $Z_3$ model with Wilson
lines described here. However, the dual heterotic model has
exactly the same superpotential, and the same light fields, hence
there are three, not just one, anomalous global $U(1)$ symmetries
also on the heterotic side. How could they appear here?
The answer is straigthforward, although somewhat unexpected.
Recall, that on the heterotic side we have also additional local
but nonanomalous $U(1)$'s. These additional factors are
spontaneously broken and their gauge bosons also decouple.
However, as discussed in specific examples, to match the spectra
we have to make certain chiral multiplets heavy on the heterotic
side, through superpotential couplings. It turns out that this
process `knocks-out' from the massless spectrum some states
charged under nonanomalous local $U(1)$'s in such a way, that what
is left are anomalous global $U(1)$'s whose anomalies and charges
are exactly the ones needed to match anomalous global factors
borne in the dual Type IIB model. This is a
somewhat unexpected observation, which might lead to
interesting phenomenological consequences. From the point of view
of the present discussion this gives further consistency check for
heterotic - type I duality in four dimensions.


\section{Conclusion}

In this paper, we have studied the properties of anomalous $U(1)$'s in a large
class of $D=4, N=1$ type IIB orientifolds, and reconsidered some candidate
evidence for heterotic-type I duality in 4 dimensions, in the light of the
recent results of Ref. \cite{Ibanez_anomalous} and \cite{Poppitz}. We have
shown that the masses of the anomalous gauge bosons are proportional
to the Planck scale, and can be made light only at the expense
of small gauge couplings, very much like in the heterotic case. They appear
to be decoupled from the Fayet-Iliopoulos terms, whose scales are set
by the values of the blowing-up modes of the underlying orbifold, and are
therefore undetermined at the perturbative level. This is a noticeable
difference with the heterotic anomalous $U(1)$, whose Fayet-Iliopoulos term
has a nonzero value, of the order of the string scale.

The absence of such Fayet-Iliopoulos terms in type IIB
orientifolds seems at first sight to contradict the generally admitted
duality between type IIB orientifolds and heterotic orbifolds, which has been
considered as a $D=4$, $N=1$ manifestation of the postulated heterotic-type I
duality in ten dimensions. However, this conclusion is too crude,
since one should compare the supersymmetric low-energy theories rather than
the original vacua. From this point of view, the vacuum shift
induced by the heterotic Fayet-Iliopoulos term appears to be a necessary,
but not always sufficient ingredient to match the gauge groups
and massless spectra of both low-energy theories. If duality holds,
one expects indeed the heterotic counterparts of the
twisted moduli that participate in the generalized Green-Schwarz mechanism
on the orientifold side to assume a vacuum expectation value in order to cancel
the heterotic Fayet-Iliopoulos term, resulting in the breaking
and decoupling at a high scale of the same number of $U(1)$'s on both
sides. At the same time, those vevs give large supersymmetric masses to
states that have no perturbative orientifold counterpart \cite{Kakushadze_Z_3},
making it possible for both massless spectra to match.

This duality picture works perfectly well for the $Z_3$ models of Ref.
\cite{Sagnotti} and \cite{Ibanez_Z_3}. However, it fails in at least two
known candidate dual examples. In the $Z_7$ model \cite{Kakushadze_Z_7}
the heterotic vacuum shift triggers the breaking of nonabelian gauge
factors, which is not required on the orientifold side. In the
$Z_3 \times Z_3$ model \cite{Kakushadze_Z_3xZ_3} some of the remaining
charged states of the heterotic model have singlet counterparts in the
orientifold model. While duality could be restored in the $Z_7$ case
if Fayet-Iliopoulos terms were generated in the orientifold model,
presumably by some nonperturbative mechanism, this would not be sufficient
in the $Z_3 \times Z_3$ case if one insists on the requirement that the
matching be enforced by the vacuum shifts. It could be that type IIB
orientifolds do not always have a heterotic dual model. This would not
necessarily contradict the conventional heterotic-type I duality
\cite{Polchinski}, since type IIB orientifolds represent a generalization
of genuine type I compactifications. Still, it would be interesting
to see how nonperturbative effects could possibly influence the notion of
heterotic-type II orientifold duality. We reserve this question for
future investigation.

Finally, let us say a few words about the possible consequences of the
orientifold anomalous $U(1)$'s. 
In heterotic string compactifications, the presence of an anomalous $U(1)$
has been shown to have numerous implications of great relevance for
phenomenology. Among them is the possibility of explaining the origin and
hierarchies of the small dimensionless parameters present in the low-energy
lagrangian, such as the Yukawa couplings \cite{fermion_masses}, in terms of
the ratio $\sqrt{|\xi^2|}/M_{Pl}\,$. Particularly encouraging is the fact that,
in explicit string models, $\xi^2$ is found to be of the order of magnitude
necessary to account for the value of the Cabibbo angle. Furthermore,
the universality of the mixed gauge anomalies implies a successful
relation between the value of the Weinberg angle at unification and the
observed fermion mass hierarchies \cite{weak_angle}.
The anomalous $U(1)$ also plays an important role in supersymmetry breaking:
not only it takes part in its mediation from the hidden sector to the
observable sector (as implied by the universal Green-Schwarz relation among
mixed gauge anomalies), but also it can trigger the breaking of
supersymmetry itself, due to an interplay between the anomalous $D$-term
and gaugino condensation \cite{susy_breaking}. Also, the heterotic anomalous
$U(1)$ is likely to have outstanding implications in cosmology, in particular
its Fayet-Iliopoulos term can dominate the vacuum energy of the early Universe,
leading to inflation \cite{inflation}. Finally, it may provide a solution of
the strong CP problem \cite{strong_CP}.

One may now ask whether the anomalous $U(1)$'s present in type IIB orientifolds
are likely to have similar consequences - or even have the potential to solve
some of the problems encountered in the heterotic case. In order to answer
this question, it is important to note that all the phenomenological
implications of the heterotic $U(1)_X$ rely on the appearance of a
Fayet-Iliopoulos term whose value, a few orders of magnitude below the Planck
mass, is fixed by the anomaly. The situation is very different in orientifolds,
where the Fayet-Iliopoulos terms are moduli-dependent: the freedom that is
gained (and allows for example to cure the problems of $D$-term inflation in
heterotic models \cite{Halyo}) is payed for by a loss of predictivity.
In that respect, one may conclude that the orientifold anomalous $U(1)$'s are
not very different from anomaly-free $U(1)$'s, whose Fayet-Iliopoulos terms
are unconstrained and can be chosen at will. 

\vskip 1cm
{\bf Acknowledgements}
\vskip .5cm
We wish to thank Jan Conrad, Ignatios Antoniadis and Luis Ib\'a\~nez for
interesting discussions and comments.
This work was partially supported by the European Commission programs
ERBFMRX-CT96-0045 and CT96-0090. Z.L. acknowledges additional financial
support from Polish Committee for Scientific Research grant 2 P03B 037 15/99.




\begin{thebibliography}{Ref}

\bibitem{GS} M. Green and J. Schwarz, Phys. Lett. B149 (1984) 117.

\bibitem{Ibanez_orientifolds} G. Aldazabal, A. Font, L.E. Ibanez,
and G. Violero, Nucl. Phys. B536 (1998) 29.

\bibitem{Sagnotti_generalized} A. Sagnotti, Phys. Lett. B294 (1992) 196.

\bibitem{Berkooz} M. Berkooz, R.G. Leigh, J. Polchinski, J.H. Schwarz,
N. Seiberg and E. Witten, Nucl. Phys. B475 (1996) 115.

\bibitem{Ibanez_anomalous} L.E. Ibanez, R. Rabadan and A.M. Uranga,
preprint FTUAM-98-16, hep-th/9808139.

\bibitem{Poppitz} E Poppitz, preprint UCSD-PTH-98-34, hep-th/9810010.

\bibitem{Polchinski} J. Polchinski and E. Witten, Nucl. Phys. B460 (1996) 525.

\bibitem{Sagnotti} C. Angelantonj, M. Bianchi, G. Pradisi, A. Sagnotti, and
Ya.S. Stanev, Phys. Lett. B385 (1996) 96.

\bibitem{Kakushadze_Z_3} Z. Kakushadze, Nucl. Phys. B512 (1998) 221.

\bibitem{Alvarez} L. Alvarez-Gaume and E. Witten, Nucl. Phys. B234 (1984) 269.

\bibitem{Fischler} W. Fischler, H.P. Nilles, J. Polchinski, S. Raby and 
L. Susskind, Phys. Rev. Lett. 47 (1981) 757.

\bibitem{Kobayashi} T. Kobayashi and H. Nakano, Nucl. Phys. B496 (1997) 103.

\bibitem{DSW} M. Dine, N. Seiberg and E. Witten, Nucl. Phys. B289 (1987) 317.

\bibitem{Atick} J. Atick, L. Dixon and A. Sen, Nucl. Phys. B292 (1987) 109;
M. Dine, I. Ichinose and N. Seiberg, Nucl. Phys. B293 (1987) 253.

\bibitem{strong_CP} J.E. Kim, Phys. Lett. B207 (1988) 434;
E.J. Chun, J.E. Kim and H.P. Nilles, Nucl. Phys. B370 (1992) 105.

\bibitem{Font} A. Font, L.E. Ibanez, H.P. Nilles and F. Quevedo,
Nucl. Phys. B307 (1988) 109; {\it ibid.}, B310 (1988) 764.

\bibitem{fermion_masses} L. Ib\'a\~nez and G. G. Ross, Phys. Lett. B332 (1994) 
100; P. Bin\'etruy and P. Ramond, Phys. Lett. B350 (1995) 49;
V. Jain and R. Shrock, Phys. Lett. B352 (1995) 83;
E. Dudas, S. Pokorski and C.A. Savoy, Phys. Lett. B356 (1995) 45;
P. Bin\'etruy, S. Lavignac,  and P. Ramond,  Nucl. Phys. B477 (1996) 353;

\bibitem{susy_breaking} P. Bin\'etruy and E. Dudas, Phys. Lett. B389 (1996)
503; G. Dvali and A. Pomarol, Phys. Rev. Lett. 77 (1996) 3728;
Z. Lalak, Nucl. Phys. B521 (1998) 37;
N. Arkani-Hamed, M. Dine, S. P. Martin, Phys. Lett. B431 (1998) 329;
T. Barreiro, B. de Carlos, J.A. Casas, J.M. Moreno,
Phys. Lett. B445 (1998) 82.


\bibitem{inflation} J.A. Casas and C.  Mu\~noz, Phys. Lett. B216 (1989) 37;
J.A. Casas, J.M. Moreno, C. Mu\~noz and M. Quiros, Nucl. Phys. B328 (1989) 272;
P. Bin\'etruy and G. Dvali, Phys. Lett. B388 (1996) 241;
E. Halyo, Phys. Lett. B387 (1996) 43.

\bibitem{orientifolds} A. Sagnotti, in Cargese `87, ``Non-Perturbative
Quantum Field Theory'', eds. G. Mack et al. (Pergamon Press, Oxford, 1988),
p. 251;
P. Horava, Nucl. Phys. B327 (1989) 461, Phys. Lett. B231 (1989) 251.

\bibitem{Ibanez_aspects}  L.E. Ibanez, C. Munoz and S. Rigolin, preprint
FTUAM-98-28, hep-ph/9812397.

\bibitem{Kakushadze_Z_7} Z. Kakushadze and G. Shiu, Phys. Rev. D56 (1997) 3686.

\bibitem{Kakushadze_Z_3xZ_3} Z. Kakushadze and G. Shiu,
Nucl. Phys. B520 (1998) 75.

\bibitem{Ibanez_Z_3} L.E. Ibanez, JHEP 9807 (1998) 002.

\bibitem{Kakushadze_orientifolds} Z. Kakushadze, G. Shiu and S.-H.H. Tye,
Nucl. Phys. B533 (1998) 25.

\bibitem{Lykken} J. Lykken, E. Poppitz and S.P. Trivedi,
preprint UCSD-PTH-98-16, hep-th/9806080.

\bibitem{Cvetic} M. Cvetic, L. Everett, P. Langacker and J. Wang,
preprint UPR-0831T, hep-th/9903051.

\bibitem{Witten} E. Witten, Phys. Lett. B149 (1984) 351.

\bibitem{weak_angle} L. Ib\'a\~nez, Phys. Lett. B303 (1993) 55;
P. Bin\'etruy and P. Ramond, Phys. Lett. B350 (1995) 49.

\bibitem{Halyo} E. Halyo, preprint SU-ITP-99-2, hep-ph/9901302. 


\end{thebibliography}
\end{document}